\documentclass[12pt]{iopart}

\usepackage{graphicx,epsfig}
\usepackage{hyperref}
\def\e{\begin{equation}}
\def\f{\end{equation}}
\def\_#1{{\bf #1}}
\def\o{\omega}

\def\.{\cdot}

\def\Im{{\rm Im\mit}}
\def\l#1{\label{eq:#1}}
\def\r#1{(\ref{eq:#1})}
\def\aee{\alpha_{\rm ee}}
\def\aem{\alpha_{\rm em}}

\def\amm{\alpha_{\rm mm}}
\def\aeeo{\alpha_{\rm ee}^{\rm co}}
\def\aeer{\alpha_{\rm ee}^{\rm cr}}
\def\aemo{\alpha_{\rm em}^{\rm co}}
\def\aemr{\alpha_{\rm em}^{\rm cr}}
\def\ameo{\alpha_{\rm me}^{\rm co}}
\def\amer{\alpha_{\rm me}^{\rm cr}}
\def\ammo{\alpha_{\rm mm}^{\rm co}}
\def\ammr{\alpha_{\rm mm}^{\rm cr}}
\def\me{\mathbf{E}_{\rm t}}
\def\mh{\mathbf{H}_{\rm t}}
\def\z{\mathbf{z}_0}

\usepackage{color}

\begin{document}

\title[]{Balanced and optimal bianisotropic particles: Maximizing power extracted from electromagnetic fields}
\author{Younes Ra'di and Sergei A. Tretyakov}

\address{Department of Radio Science and Engineering/SMARAD Center of Excellence, Aalto University, P.~O.~Box~13000, FI-00076 AALTO, Finland}
\ead{younes.radi@aalto.fi}
\begin{abstract}
Here we introduce the concept of ``optimal
particles" for strong interactions with electromagnetic fields. We assume that a particle occupies a given electrically small volume in space
and study the required optimal relations between the particle polarizabilities. In these optimal
particles, the inclusion shape and material are chosen so that the particles extract
the maximum possible power from given incident  fields.   It appears that for different excitation scenarios the
optimal particles are bianisotropic chiral, omega, moving, and Tellegen particles. The optimal dimensions  of resonant canonical chiral and omega particles are found analytically. Such optimal particles have extreme properties in scattering (for example, zero backscattering or invisibility). Planar arrays of optimal particles possess extreme properties in reflection and transmission (e.g., total absorption or magnetic-wall response), and volumetric composites of optimal particles realize, for example, such extreme materials as the chiral nihility medium.

\end{abstract}

\maketitle

\section{Introduction}
Let us consider an electrically small particle excited by a given
external electromagnetic field, for example, a plane wave. We assume
that the particle is electrically small and its response can be
modelled in the dipole (electric and magnetic) approximation. The
incident fields polarize the particle, and we are interested in the
question how this effect can be maximized. Of course, simply
increasing the values of the particle polarizabilities (for example,
increasing the particle size or bringing it to a resonance), we
increase the amplitudes of the excited dipole moments. But what
would be the optimal relation between the polarizabilities of the
particle to ensure the most effective interaction with the incident
fields, provided that the overall size of the particle is fixed and
we cannot increase the absolute values of the polarizabilities?
In this study we consider particles which respond only as
electric and magnetic dipoles. If arbitrarily high-order multipole
moments are resonantly excited and contribute to the process of
receiving power, there is no limit on how much power can be absorbed
by a finite-size particle. This follows from the general relation
between the maximal absorption cross section $\sigma_{\rm abs\,\,
max}$ and gain $G$ of an arbitrary reciprocal antenna operating in a
reciprocal environment (see, e.g. \cite[eq.~(2.111)]{Balanis})  \e
S_{\rm abs\,\, max}=G\, {\lambda^2\over 4\pi}\l{AG} \f where
$\lambda$ is the wavelength. This clearly tells the (obvious) fact
that received power can be increased by making the antenna
directive, and it is known that in the assumption of perfect
radiation efficiency (no losses in the antenna body), the antenna
gain has no upper limit for a fixed antenna size \cite{Bow}.
However, increasing directivity without increasing antenna size
(realizing super-directive antennas or particles) implies sharp
sub-wavelength variations of the antenna current, which leads to
practical limitations. In microwave region, super-directive antennas
can be realized as antenna arrays (e.g. \cite{Salt}), while in the
optical region standing surface plasmon-polariton oscillations can
be used \cite{Luo,Ng,Pendry}.

Here we focus on maximizing the extinction cross section
(formula \r{AG} applies to the absorption cross section) of general
dipole particles. Recently, it was noticed that the shape of chiral
particles (for example, metal helices) can be optimized in such a
way that these optimal helices radiate waves of only one circular
polarization, whatever is the polarization of the exciting field
\cite{gent,russian_2009}. Later, it was shown that the optimal
spiral shape corresponds to the maximum (or minimum, equal zero)
reactive energy of the particle in a given plane-wave field
\cite{morocco}, interacting with the wave in the most effective way.
It has been therefore established that there exist particles (and
metamaterials formed by these particles), whose properties are
optimal for strong interactions electromagnetic fields. Furthermore,
mixtures of optimal spirals realize effective media with the index
of refraction $n=-1$ for one of the circular polarizations, while
the same medium is transparent for the orthogonal circular
polarization \cite{unit}. Analytical and numerical studies of
optimal helical particles have been done in \cite{Saenz}. However,
optimal helices show their optimal and extreme properties only if
excited by propagating circularly polarized plane waves. First steps
towards generalizations for other exciting fields have been done in
\cite{Karlsruhe,Belarus}, where the case of linearly polarized
evanescent fields has been discussed.

The goal of the present work is to find the optimal particles
which would interact most effectively with waves of other polarizations (in particular, linear) and also
with both propagating and evanescent incident fields.
We expect that such optimal inclusions will open a way to
realizing particle arrays and volumetric composites with extreme properties, as we
see from the known example of the optimal spiral.

The strength of field-particle interactions can be measured in various ways. Here we
define the optimal particle as one which extracts the maximum (or minimum) power from given incident fields.

\section{General framework}
From the known example of the optimal chiral particle it is clear that in order to enhance the field-particle
interaction, we need to design a particle which would be polarizable by
both electric and magnetic fields, and the particle should be
bianisotropic, and not just an electrically and
magnetically polarizable inclusion. In the presence of magneto-electric coupling, excitations of the
particle due to the incident electric and magnetic fields can sum up coherently, maximizing the total excitation.
Thus, we assume that the particle for optimally strong interactions with different kinds of waves is
a bianisotropic particle whose electric and magnetic moments are
induced by both electric and magnetic fields $\_E$ and $\_H$. The general
polarizability relation can be written as
\begin{equation}
\left[ \begin{array}{c} \mathbf{p} \\ \mathbf{m}\end{array} \right]
=\left[ \begin{array}{cc} \overline{\overline{\alpha}}_{\rm ee} & \overline{\overline{\alpha}}_{\rm em}
 \\  \overline{\overline{\alpha}}_{\rm me} & \overline{\overline{\alpha}}_{\rm mm} \end{array}
\right]\cdot \left[ \begin{array}{c} \mathbf{E} \\ \mathbf{H}\end{array} \right] ,
\label{eq:a}\end{equation}
where $\mathbf{p}$ and $\mathbf{m}$
are the induced electric and magnetic dipole moments, respectively,
and $\overline{\overline{\alpha}}_{\rm ij}$ are the polarizability dyadics.

Let us define a direction is space (along the unit vector $\_z_0$) and assume that the particle is excited by a plane wave propagating along this direction. In other scenarios, we can assume that the particles form a planar array in the plane orthogonal to $\_z_0$, or that the particles form a 3D periodical crystal forming a half space (or a finite-thickness layer), with the surface orthogonal to $\_z_0$. In all these scenarios, the most interesting case is the case of the uniaxial symmetry of the particles and the polarizability dyadics. To simplify the following derivations, we will in addition assume that the particle is not excited by the fields directed along the axis $\_z_0$. This fully covers the case of excitation by plane waves propagating along the axis $\_z_0$ and the case of the normal incidence on planar arrays and planar interfaces. For the general excitation, the assumption means that the theory is restricted to ``planar'' particles, which feel only the fields in the plane, orthogonal to $\_z_0$. Thus, we write the polarizability dyadics of uniaxial bianisotropic particles in the form
\begin{equation}
\begin{array}{l}\overline{\overline{\alpha}}_{\rm ee} =\aeeo \overline{\overline{I}}_{t}+\aeer \overline{\overline{J}}_{t},\\\displaystyle
\overline{\overline{\alpha}}_{\rm mm} =\ammo \overline{\overline{I}}_{t}+\ammr \overline{\overline{J}}_{t},\\\vspace*{.1cm}\displaystyle
\overline{\overline{\alpha}}_{\rm em} =\aemo \overline{\overline{I}}_{t}+\aemr \overline{\overline{J}}_{t},\\\vspace*{.1cm}\displaystyle
\overline{\overline{\alpha}}_{\rm me} =\ameo \overline{\overline{I}}_{t}+\amer \overline{\overline{J}}_{t}.
\end{array}\label{eq:b}
\end{equation}
Here, $\overline{\overline{I}}_t=\overline{\overline{I}}-\mathbf{z}_0\mathbf{z}_0$
is the transverse unit dyadic, $\overline{\overline{I}}$ is the 3D unit dyadics, and
$\overline{\overline{J}}_t=\mathbf{z}_0\times\overline{\overline{I}}_t$
is the vector-product operator.

In the following, classification of bianisotropic particles in terms of reciprocity and the symmetry of their magneto-electric coefficients will be important. We will discuss the two reciprocal classes (chiral and omega) and two nonreciprocal classes (moving and Tellegen) \cite{basic}. For reciprocal particles the electric and magnetic polarizabilities are always symmetric dyadics, so we keep only $\aeeo$ and $\ammo$ in the expressions for $\overline{\overline{\alpha}}_{\rm ee}$ and  $\overline{\overline{\alpha}}_{\rm mm}$. For the nonreciprocal classes we keep also the anti-symmetric parts $\aeer$ and $\ammr$ of these dyadics. The polarizability dyadics of particles of these four classes are summarized in Table~\ref{ta:main-classes}.

\begin{table}[h]
\caption{Uniaxial bianisotropic particles}
\begin{tabular}{|p{35mm}|p{39mm}|p{43mm}|p{43mm}|}
\hline
\multicolumn{4}{|l|}{{\bf Main classes of uniaxial bianisotropic particles}}
\\
\hline
Omega
&
Chiral
&
Moving
&
Tellegen
 \\
\hline
\vspace{0.5mm} $\begin{array}{c}
 \overline{\overline{\alpha}}_{\rm ee}=\aeeo \overline{\overline{I}}_{t}\\\displaystyle
\overline{\overline{\alpha}}_{\rm mm}=\ammo \overline{\overline{I}}_{t}\\\vspace*{.1cm}\displaystyle
\overline{\overline{\alpha}}_{\rm em}=\overline{\overline{\alpha}}_{\rm me}=j\Omega \overline{\overline{J}}_{t}\\
\end{array}$
&
\vspace{0.5mm} $\begin{array}{c}
 \overline{\overline{\alpha}}_{\rm ee}=\aeeo\overline{\overline{I}}_{t}\\\displaystyle
\overline{\overline{\alpha}}_{\rm mm}=\ammo\overline{\overline{I}}_{t}\\\vspace*{.1cm}\displaystyle
\overline{\overline{\alpha}}_{\rm em}=-\overline{\overline{\alpha}}_{\rm me}=j\kappa\overline{\overline{I}}_{t}\\
\end{array}$

&
\vspace{0.5mm}$\begin{array}{c}\overline{\overline{\alpha}}_{\rm ee}=\aeeo\overline{\overline{I}}_{t}+j\aeer\overline{\overline{J}}_{t}\\\displaystyle
\overline{\overline{\alpha}}_{\rm mm}=\ammo\overline{\overline{I}}_{t}+j\ammr\overline{\overline{J}}_{t}\\\vspace*{.1cm}\displaystyle
\overline{\overline{\alpha}}_{\rm em}=-\overline{\overline{\alpha}}_{\rm me}=V\overline{\overline{J}}_{t}\\
\end{array}$

&
\vspace{0.5mm}$\begin{array}{c}\overline{\overline{\alpha}}_{\rm ee}=\aeeo\overline{\overline{I}}_{t}+j\aeer\overline{\overline{J}}_{t}\\\displaystyle
\overline{\overline{\alpha}}_{\rm mm}=\ammo\overline{\overline{I}}_{t}+j\ammr\overline{\overline{J}}_{t}\\\vspace*{.1cm}\displaystyle
\overline{\overline{\alpha}}_{\rm em}=\overline{\overline{\alpha}}_{\rm me}=\chi\overline{\overline{I}}_{t}\\
\end{array}$
\\
\hline

\end{tabular}
\label{ta:main-classes}
\end{table}

\noindent
The imaginary units in the notations for the coupling parameters of reciprocal particles and in the anti-symmetric parts of electric and magnetic polarizabilities are introduced in order to ensure that the absorption and scattering losses are measured by the imaginary parts of all the polarizability components. This follows from the general conditions for lossless particles (e.g. \cite{basic}):
\begin{equation}
\begin{array}{l}
\overline{\overline{\alpha}}_{\rm ee} = \overline{\overline{\alpha}}_{\rm ee}^\dag,\vspace*{.2cm}\\\displaystyle
\overline{\overline{\alpha}}_{\rm mm} = \overline{\overline{\alpha}}_{\rm mm}^\dag,\vspace*{.2cm}\\\displaystyle
\overline{\overline{\alpha}}_{\rm em} = \overline{\overline{\alpha}}_{\rm me}^\dag,
\end{array}
\label{eq:c}
\end{equation}
where  $\dag$ is the Hermitian conjugate operator (transpose of the complex conjugate).

Next, we specify the assumptions about incident electromagnetic fields. We will study excitations in form of plane waves (propagating or evanescent) or their combinations (standing waves).
As the particle reacts on the incident electric and magnetic fields in the $\mathbf{x}-\mathbf{y}$ plane, we separate the fields of a general plane wave in free space into transverse and normal field components:
\begin{equation}
\begin{array}{l}
\mathbf{H}=\mh+H_z\mathbf{z}_0,\vspace*{.2cm}\\\displaystyle
\mathbf{E}=\me+E_z\mathbf{z}_0 ,
\end{array}
\label{eq:d}\end{equation}
where
\begin{equation}
\begin{array}{l}
\mathbf{H}_{\rm t}=H_x\mathbf{x}_0+H_y\mathbf{y}_0,\vspace*{.2cm}\\\displaystyle
\mathbf{E}_{\rm t}=-\overline{\overline{Z}}\cdot \mathbf{z}_0\times\mathbf{H}_{\rm t}.
\end{array}
\label{eq:d1}\end{equation}
The wave vector $\_k$ of a plane wave we also split into the longitudinal part $\beta\_z_0$ and the transverse part $\_k_{\rm t}$: $\_k=\_k_{\rm t}+\beta\_z_0$, where $\beta=(\displaystyle\omega^2 \mu_0 \epsilon_0-k_t^2)^{1/2}$. The dependence on the  coordinates is $\exp(-j\mathbf{k}\cdot \mathbf{r})$, and the harmonic time dependence is in the form $\exp (j\omega t)$. The wave impedance relates the transverse components of the fields, and it can be written as
\begin{equation}
\begin{array}{l}
\displaystyle
\overline{\overline{Z}}=Z_{\rm TM}\frac{\mathbf{k}_{\rm t} \mathbf{k}_{\rm t}}{k_{\rm t}^2}+
Z_{\rm TE}\frac{\mathbf{k}_{\rm t} \times \mathbf{z}_0   \mathbf{k}_{\rm t} \times \mathbf{z}_0}{k_{\rm t}^2},
\vspace*{.2cm}\\\displaystyle
Z_{\rm TM}=\frac{\beta}{\omega\epsilon_0},\vspace*{.2cm}\\\displaystyle
Z_{\rm TE}=\frac{\omega\mu_0}{\beta}.
\label{eq:e}\end{array}
\end{equation}
Here, $Z_{\rm TM}$ and $Z_{\rm TE}$ are the characteristic impedances for the two linear eigenpolarizations (transverse electric, TE, and transverse magnetic, TM, waves are defined with respect to the vector $\_z_0$).

Next, we write the general expression for the time-averaged power extracted by the particle from the incident field (the external field power spent on the excitation of the particle):
\begin{equation}
\begin{array}{l}
\displaystyle
P=-{\omega\over 2}{\rm Im}\{\_p\cdot \me^*+ \_m\cdot \_H_{\rm t}^*\},
\label{eq:f}\end{array}
\end{equation}
where $*$ denotes the complex conjugate operator. Using (\ref{eq:a}) and (\ref{eq:b}), the electric and magnetic moments of bianisotropic particles can be generally written as
\begin{equation}
\begin{array}{l}\displaystyle
\_p=\aeeo\me+\aeer\z\times\me+\aemo\mh+\aemr\z\times\mh,\vspace*{.2cm}\\\displaystyle
\_m=\ameo\me+\amer\z\times\me+\ammo\mh+\ammr\z\times\mh .
\end{array} \label{eq:g}\end{equation}
We express the transverse components of the incident field in the Cartesian coordinates $x$, $y$, and substitute these moments in (\ref{eq:f}). The result reads
\begin{equation}
\begin{array}{l}
\displaystyle
P=-{\omega\over 2}{\rm Im}\left\{\aeeo|\me|^2\right.
\vspace*{.2cm}\\\displaystyle\hspace*{2.9cm}
+2\aeer\Im\left(E_x^* E_y\right)
\vspace*{.2cm}\\\displaystyle\hspace*{2.9cm}
 +\aemo\left(-\frac{1}{Z_{\rm TE}}E_x^* E_y+\frac{1}{Z_{\rm TM}}E_x E_y^*\right)
\vspace*{.2cm}\\\displaystyle\hspace*{2.9cm}
+\aemr\left(-\frac{1}{Z_{\rm TE}}|E_y|^2-\frac{1}{Z_{\rm TM}}|E_x|^2\right)
\vspace*{.2cm}\\\displaystyle\hspace*{2.9cm}
+\ameo\left(-\frac{1}{Z_{\rm TE}^*}E_x E_y^*+\frac{1}{Z_{\rm TM}^*}E_x^* E_y\right)
\vspace*{.2cm}\\\displaystyle\hspace*{2.9cm}
+\amer\left(\frac{1}{Z_{\rm TE}^*}|E_y|^2+\frac{1}{Z_{\rm TM}^*}|E_x|^2\right)
\vspace*{.2cm}\\\displaystyle\hspace*{2.9cm}
+\ammo\left(\frac{1}{|Z_{\rm TE}|^2}|E_y|^2+\frac{1}{|Z_{\rm TM}|^2}|E_x|^2\right)
\vspace*{.2cm}\\\displaystyle\hspace*{2.9cm}
\left. +2\ammr\Im\left(\frac{1}{Z_{\rm TE}Z_{\rm TM}^*}E_x^{*} E_y\right)\right\}.
\label{eq:h}\end{array}
\end{equation}
This is the power extracted by a bianisotropic particle  from
a given arbitrary polarized field. In what follows, we will find
the optimal relations between the particle polarizabilities which
maximize this value, providing recipes for the design of particles
for extremely strong interactions with electromagnetic fields.

\section{Optimal particle classes}
\label{classes}

Let us assume that the amplitudes of the polarizabilities in (\ref{eq:h}) are limited or fixed by the particle size and material. In this situation, it is clear that the topology of the optimal particle should be chosen so that the terms containing the magneto-electric coefficients would contribute constructively to the total extracted power. This consideration will determine
the optimal particle classes for different types of excitation, because the impedance values in (\ref{eq:h}) can take real or imaginary values for propagating waves or reactive external fields.

\subsection{Excitation by linearly polarized propagating waves}

If the particle is excited by a linearly polarized propagating plane
wave, then the two transverse field components $E_x$ and $E_y$ are
in phase or out of phase ($E_x E_y^*$ is a real number), and the
wave impedances $Z_{\rm TE}$ and $Z_{\rm TM}$ are real numbers.  The
expression for the received power in (\ref{eq:h}) takes the
form
\begin{equation}
\begin{array}{l}
\displaystyle
P=-{\omega\over 2}|\me|^2{\rm Im}\left\{\aeeo\right.
\vspace*{.2cm}\\\displaystyle\hspace*{2.9cm}
 +2\chi\, \left(-\frac{1}{Z_{\rm TE}}\frac{E_x^* E_y}{|\me|^2}+\frac{1}{Z_{\rm TM}}\frac{E_x E_y^*}{|\me|^2}\right)
\vspace*{.2cm}\\\displaystyle\hspace*{2.9cm}
-2V\, \left(\frac{1}{Z_{\rm TE}}\frac{|E_y|^2}{|\me|^2}+\frac{1}{Z_{\rm TM}}\frac{|E_x|^2}{|\me|^2}\right)
\vspace*{.2cm}\\\displaystyle\hspace*{2.9cm}\left.
+\ammo\left(\frac{1}{|Z_{\rm TE}|^2}\frac{|E_y|^2}{|\me|^2}+\frac{1}{|Z_{\rm TM}|^2}\frac{|E_x|^2}{|\me|^2}\right)\right\}.
\label{eq:i}\end{array}
\end{equation}
The terms related to omega and chiral coupling do not contribute to
(\ref{eq:i}) and the remaining co- and cross-coupling
polarizabilities refer to the Tellegen and moving particles,
respectively. It means  that for receiving the maximally
possible power from incident propagating linearly polarized waves,
particles with nonreciprocal magneto-electric coupling are required
(both Tellegen and ``moving'' coupling mechanisms). In the case of
an axially incident wave (propagating along $\_z_0$) both impedances
are equal to $\eta_0=\sqrt{\mu_0\over \epsilon_0}$, the term
concerning to the Tellegen coupling cancels out, and the received
power can be written as
\begin{equation}
\begin{array}{l}
\displaystyle
P=-{\omega\over 2}|\me|^2{\rm Im}\left\{\aeeo-\frac{2}{\eta_0}V+\frac{1}{\eta_0^2}\ammo\right\}.
\label{eq:j}\end{array}
\end{equation}
Therefore, in the case of an axially incident linearly polarized propagating wave, the optimal particle class is that of moving particles (nonreciprocal anti-symmetric magneto-electric coupling).

\subsection{Excitation by linearly polarized reactive fields}

In case of reactive linearly polarized field in free space (for
example, an evanescent plane wave or a standing linearly polarized
wave) the transverse electric and magnetic fields are $90^\circ$ out
of phase and both impedances $Z_{\rm TE}$ and $Z_{\rm TM}$ are
purely imaginary.  The received power in (\ref{eq:h}) can be
written as
\begin{equation}
\begin{array}{l}
\displaystyle
P=-{\omega\over 2}|\me|^2{\rm Im}\left\{\aeeo\right.
\vspace*{.2cm}\\\displaystyle\hspace*{2.9cm}
 -2\kappa\, \Im\left\{-\frac{1}{Z_{\rm TE}}\frac{E_x^* E_y}{|\me|^2}+\frac{1}{Z_{\rm TM}}\frac{E_x E_y^*}{|\me|^2}\right\}
\vspace*{.2cm}\\\displaystyle\hspace*{2.9cm}
+2\Omega\, \Im\left\{\frac{1}{Z_{\rm TE}}\frac{|E_y|^2}{|\me|^2}+\frac{1}{Z_{\rm TM}}\frac{|E_x|^2}{|\me|^2}\right\}
\vspace*{.2cm}\\\displaystyle\hspace*{2.9cm}\left.
+\ammo\left(\frac{1}{|Z_{\rm TE}|^2}\frac{|E_y|^2}{|\me|^2}+\frac{1}{|Z_{\rm TM}|^2}\frac{|E_x|^2}{|\me|^2}\right)\right\} .
\label{eq:k}\end{array}
\end{equation}
We see that in this case the received power can be enhanced
introducing reciprocal chiral and omega coupling effects. If the two
impedances are equal $Z_{\rm TE}=Z_{\rm TM}=Z$ (for example, a
standing plane wave at the plane equally distanced from the maxima
of the two fields), the term with the chirality parameter cancels
out and we can write an expression analogous to that for the
propagating waves (\ref{eq:j}):
\begin{equation}
\begin{array}{l}
\displaystyle
P=-{\omega\over 2}|\me|^2{\rm Im}\left\{\aeeo+2\Omega{\rm Im}\left(\frac{1}{Z}\right)+\frac{1}{|Z|^2}\ammo\right\}.
\label{eq:j_evan}\end{array}
\end{equation}

We can observe that for optimizing interactions with propagating linearly polarized waves, nonreciprocal particles are needed, whereas in the case of linearly polarized reactive fields, we need reciprocal (omega) particles. It is interesting to note that the omega particle was introduced as a particle strongly interacting with linearly polarized fields \cite{Engheta-omega,sochava}, but it was noticed that they are optimal for linearly polarized evanescent fields.

\subsection{Excitation by circularly polarized propagating waves}

In the general case, we define the circularly polarized incident field as a field configuration where the electric field vector is circularly polarized in the transverse plane of the particle, that is, $E_y=\pm j E_x$ ($E_x E_y^*$ is an imaginary number). For circularly polarized waves, the upper and lower signs correspond to left- and right-hand circularly polarized waves propagating along $\_z_0$, respectively. In general, magnetic field can be elliptically polarized.

For propagating waves, the impedances $Z_{\rm TE,TM}$ are real numbers, and the power extracted by the particle according to  (\ref{eq:h}) can be written as
\begin{equation}
\begin{array}{l}
\displaystyle
P=-{\omega\over 2}|\me|^2{\rm Im}\left\{\aeeo\right.
\vspace*{.2cm}\\\displaystyle\hspace*{2.9cm}
 -\left(V\mp \kappa \right)\left(\frac{1}{Z_{\rm TE}}+\frac{1}{Z_{\rm TM}}\right)
\vspace*{.2cm}\\\displaystyle\hspace*{2.9cm}
+\frac{1}{2}\ammo\left(\frac{1}{|Z_{\rm TE}|^2}+\frac{1}{|Z_{\rm TM}|^2}\right)
\vspace*{.2cm}\\\displaystyle\hspace*{2.9cm}
\left. \pm \left(\aeer+\frac{\ammr}{\eta_0^2}\right)\right\}.
\label{eq:l}\end{array}
\end{equation}
Here, the contributing co- and cross-coupling polarizabilities stand for reciprocal (chiral) and nonreciprocal (moving) classes, and the terms concerning to omega and Tellegen properties do not have any role in interacting with circularly polarized propagating waves. In addition, the antisymmetric parts of electric and magnetic polarizabilities contribute to the received power. As it can be seen from (\ref{eq:l}), in the case of axially incident waves, both chiral and moving-particle coupling terms are essential, and the power can be written as
\begin{equation}
\begin{array}{l}
\displaystyle
P=-{\omega\over 2}|\me|^2{\rm Im}\left\{\aeeo-\frac{2}{\eta_0}\left(V\mp \kappa\right)
+\frac{1}{\eta_0^2}\ammo\pm \left(\aeer+\frac{\ammr}{\eta_0^2}\right)\right\}.
\label{eq:m}\end{array}
\end{equation}
This is in agreement with the earlier studies of chiral particles as the optimal particles for interactions with propagating circularly polarized waves \cite{gent,morocco,unit}, where it was assumed that the particles were reciprocal (that is, parameters $V$, $\aeer$ and $\ammr$ equal zero).

\subsection{Excitation by circularly polarized evanescent fields}

In this case, we assume  that the wave impedances are purely
imaginary and find that the received power in (\ref{eq:h})
can be simplified to
\begin{equation}
\begin{array}{l}
\displaystyle
P=-{\omega\over 2}|\me|^2{\rm Im}\left\{\aeeo\right.
\vspace*{.2cm}\\\displaystyle\hspace*{2.9cm}
+\left(\Omega\pm \chi\right)\Im\left\{\frac{1}{Z_{\rm TE}}+\frac{1}{Z_{\rm TM}}\right\}
\vspace*{.2cm}\\\displaystyle\hspace*{2.9cm}
+\frac{1}{2}\ammo\left(\frac{1}{|Z_{\rm TE}|^2}+\frac{1}{|Z_{\rm TM}|^2}\right)
\vspace*{.2cm}\\\displaystyle\hspace*{2.9cm}
\left. \pm \left(\aeer-\frac{\ammr}{\eta_0^2}\right)\right\}.
\label{eq:n}\end{array}
\end{equation}
Similarly to the case of propagating circularly polarized waves, both reciprocal and nonreciprocal magneto-electric coupling contribute to the extracted power, but here we need the other two classes of magneto-electric coupling: reciprocal omega coupling and nonreciprocal Tellegen coupling.

\section{Optimal reciprocal particles}
\label{orp}

In the previous section, we found what classes of magneto-electric particles are most suitable for interactions with different types of exciting fields (propagating versus evanescent, linearly versus circularly polarized). The next step is to find the optimal particle shapes for these particles. In this section, we will do it for reciprocal particles: chiral and omega particles, optimal for circularly polarized propagating waves and evanescent linearly polarized fields. Optimal chiral particles have been studied earlier \cite{gent,morocco,russian_2009,unit,Saenz}, but in those papers the adopted criteria for the optimal field-particle interaction were different: Earlier introduced optimal helices react only on circularly polarized waves of one of the two orthogonal polarizations and store the maximum reactive field defined in the quasi-static approximation. With this in view, we will compare our results for the optimal dimensions of helices optimized in the sense of the maximum extracted power from the incident fields.

The optimal shape of electrically small reciprocal particles is possible to determine under quite general assumptions thanks to the existence of so called ``hierarchy of
polarizabilities". Considering the particle size ($d$) normalized to the free-space wavelength $d/\lambda$ as a small parameter, it is known that the
electric polarizability is the strongest (exists even in the zero-frequency limit), then comes the
cross-coupling coefficient (a first-order spatial dispersion effect of the order of $d/\lambda$), and the
weakest effect is the artificial magnetism (a second-order effect of the order of $(d/\lambda)^2$) \cite{hie1,hie2,hie3,basic,Amodel}. To estimate the polarizabilities and find the optimal particle shape analytically we will use the antenna model for thin-wire scatterers (chiral and omega) \cite{Amodel}. We will see that the solution for the optimal particle shape will define an optimal value for the ``scaling parameter'' for the polarizabilities of reciprocal particles.

\subsection{Chiral particles}

\begin{figure}[h!]
\centering
\epsfig{file=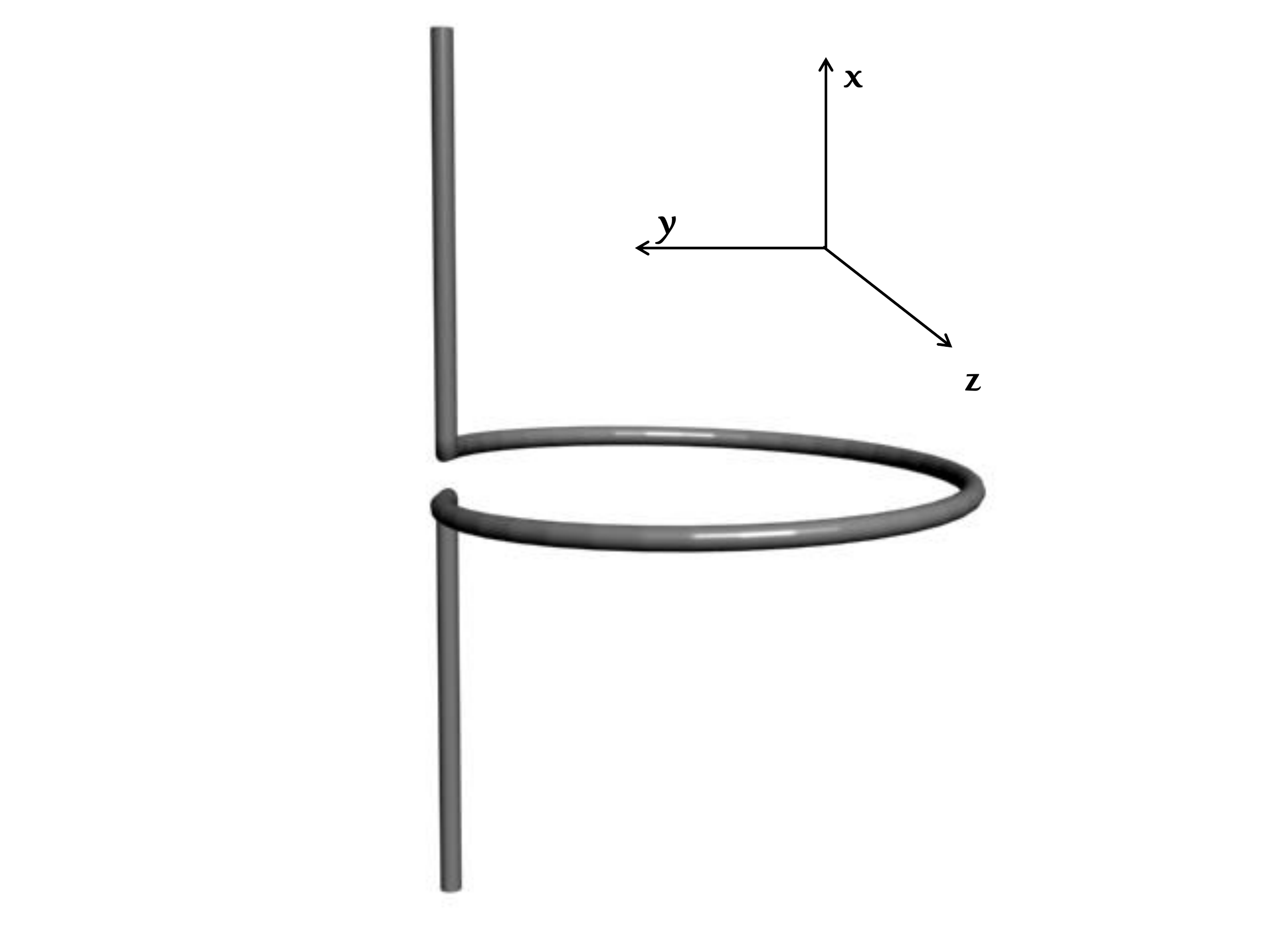, width=0.4\textwidth}
\caption{Canonical chiral particle. A uniaxial particle with the polarizabilities defined in Table~\ref{ta:main-classes} can be formed by adding a second identical particle rotated by $90^\circ$ around the axis $z$.}
\label{chiral_shape}
\end{figure}

According to the above results, chiral particles can be optimized for interactions with propagating circularly polarized waves. As a particular example of a small chiral particle, interacting with fields in the microwave region, we will consider the well-known metal canonical particle shown in Figure~\ref{chiral_shape}.
This particle can be viewed as a connection of two small wire antennas: a short electric dipole antenna of length $2l\ll\lambda$ and a small loop antenna with the radius $a\ll\lambda$ \cite{Amodel}. The plane of the loop is orthogonal to the electric dipole antenna. This simple shape of a metal helix was probably introduced for the first time by D. Jaggard et al. \cite{Jaggard} and then used in a number of studies as a canonical example of a chiral particle which allows simple analytical solutions for its polarizabilities (see e.g. \cite{basic,Amodel} and references therein).

In terms of the parameters of the two small antennas forming the particle, the polarizabilities along axis $x$ can be written as \cite{Amodel}
\begin{equation}
\begin{array}{l}
\displaystyle
\aee=-j\frac{l^2}{\omega Z_{\rm tot}},\vspace*{.3cm}\\\displaystyle
\aem=j\kappa =\pm j \aee\eta_0\frac{k_0 S}{l},\vspace*{.3cm}\\\displaystyle
\amm=\aee\left(\eta_0\frac{k_0S}{l}\right)^2.
\end{array}\label{eq:aa}\end{equation}
Here, $Z_{\rm tot}$ is the total impedance of the
particle (the sum of the input impedances of the loop and  electric dipole antennas), $S=\pi a^2$ is the loop area, and $k_0=\omega\sqrt{\epsilon_0\mu_0}$ is the free-space wave number. At the resonance of the particle, $Z_{\rm tot}=R_{\rm
tot}+jX_{\rm tot}$ is a real number, which is non-zero because of inevitably present scattering losses even in the absence of absorption in the particle. Thus, at the resonance frequency, the electric and magnetic polarizabilities $\alpha_{\rm ee}$, $\alpha_{\rm mm}$,  and the coupling (chirality) parameter $\kappa$ are imaginary and,  as it can be seen from (\ref{eq:l}), all of them contribute to absorption of field energy.

The received power (\ref{eq:l}) can be written as
\begin{equation}
\begin{array}{l}
\displaystyle
P=-{\omega\over 2}|\me|^2{\rm Im}\left\{\aeeo\pm \kappa \left(\frac{1}{Z_{\rm TE}}+\frac{1}{Z_{\rm TM}}\right)
+\frac{1}{2}\ammo\left(\frac{1}{|Z_{\rm TE}|^2}+\frac{1}{|Z_{\rm TM}|^2}\right)\right\}.
\label{eq:bb}\end{array}
\end{equation}
Substituting the characteristic impedances from (\ref{eq:e}), we get
the received power at the resonance frequency as
\begin{equation}
\begin{array}{l}
\displaystyle
P={l^2\over 2}|\me|^2 \frac{1}{R_{\rm tot}}\left\{1+\frac{k_0S}{l}\left(\frac{\beta}{k_0}+\frac{k_0}{\beta}\right)
+\frac{1}{2}\left(\frac{k_0S}{l}\right)^2\left(\frac{\beta^2}{k_0^2}+\frac{k_0^2}{\beta^2}\right)\right\}.
\label{eq:cc}\end{array}
\end{equation}
We simplify our problem assuming that the incidence direction is
orthogonal to the  induced electric and magnetic moments (along the
axis $z$ of Figure~\ref{chiral_shape}), which means $\beta=k_0$. The
received power simplifies to
\begin{equation}
\begin{array}{l}
\displaystyle
P={l^2\over 2}|\me|^2 \frac{1}{R_{\rm tot}}\left\{1+ 2\frac{k_0S}{l}
+\left(\frac{k_0S}{l}\right)^2\right\}.
\label{eq:dd}\end{array}
\end{equation}
For particles without absorption loss, $R_{\rm tot}$ is only due to the scattering loss,
and its value is known from the theory of small antennas (e.g., \cite{Schelk,LoLee}):
\begin{equation}
\displaystyle
R_{\rm tot}=\frac{\eta_0}{6\pi}\left(k_0^2l^2+k_0^4S^2\right).
\label{eq:v}\end{equation}
Substituting this value, after some simple algebra, we find $P$ as
\begin{equation}
\begin{array}{l}
\displaystyle
P={|\me|^2\over 2} \frac{6\pi}{\eta_0 k_0^2}\left\{1+\frac{\displaystyle \frac{4l}{\lambda}  \left({\lambda\over 4}-l\right)^2}{\displaystyle l^2+\frac{4}{\lambda^2}\left({\lambda\over 4}-l\right)^4}\right\}.
\label{eq:ee}\end{array}
\end{equation}
Here we have used the resonance condition for the particle: at the fundamental resonance frequency the total (stretched) length $2l+2\pi a$ of the wire is approximately $\lambda/2$.
To determine the optimal relation between $l$ and $a$ which would bring the received power $P$ to an extremum, we calculate its derivative with respect to $l$ and equate it to zero under the condition of resonance. This determines the optimal dimensions:
\begin{equation}
\begin{array}{l}
\displaystyle
l_{\rm opt}=(2-\sqrt{3})\frac{\lambda}{4},\vspace*{.3cm}\\\displaystyle
a_{\rm opt}=\frac{\sqrt{3}-1}{\pi}\frac{\lambda}{4}.
\label{eq:ff}\end{array}
\end{equation}
This corresponds to the optimal value of the  ``scaling factor"
$k_0S/l=1$ in (\ref{eq:dd}), which means that all the
normalized polarizabilities of optimal chiral particles are equal:
\begin{equation}
\eta_0\alpha_{\rm ee}=\mp \kappa=\frac{1}{\eta_0}\alpha_{\rm mm}.
\label{eq:gg}\end{equation}

This last result coincides with that in \cite{gent,morocco,unit} and the optimal dimensions (\ref{eq:ff}) are the same as found in \cite{Saenz}, but an all these earlier papers the optimal particles were defined as those interacting only with circularly polarized waves of one of the two eigenpolarizations. Now, we see that these particles are also optimal in the sense of the maximum received power.
The physical nature of this optimal behaviour of helices is simple: In
one extreme case the electromotive forces induced by electric and
magnetic incident fields cancel out, and the total induced current is zero,
while in the other extreme case the two forces sum up in phase and
interaction is optimally enhanced.
Thus, we see that all the known definitions of optimal chiral particles lead to the same solution for the particle shape and to the same general balance relation between the polarizabilities (\ref{eq:gg}), if the incident wave propagates along the particle axis.
Note that in general, the optimal particle dimensions
 depend not only on the
frequency, but also on the transverse wave number (the incidence direction) of
the exciting plane wave, as is seen from (\ref{eq:cc}).

\subsection{Omega particles}

\begin{figure}[h!]
\centering
\epsfig{file=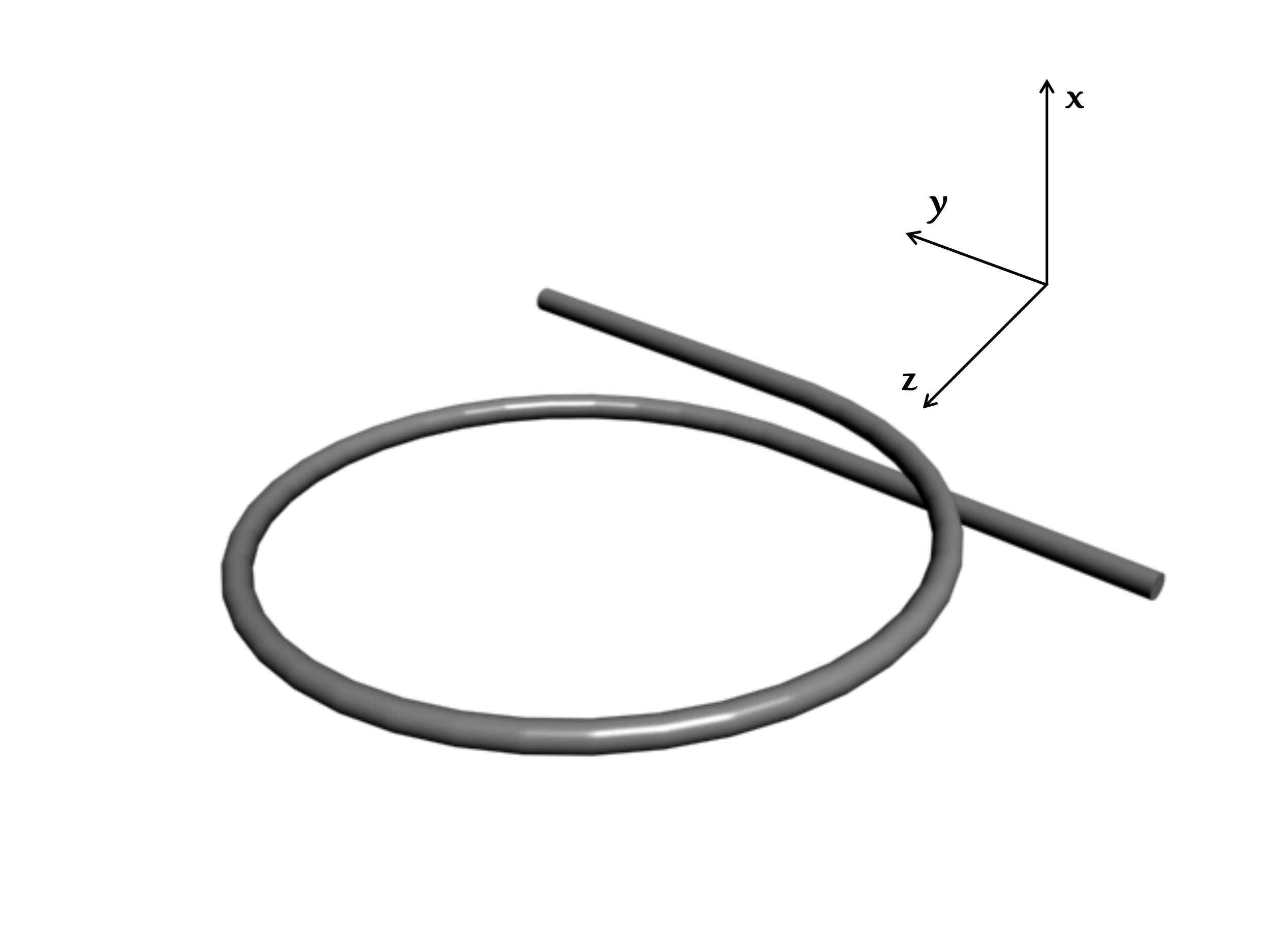, width=0.4\textwidth}
\caption{Canonical omega particle. A uniaxial particle with the polarizabilities defined in Table~\ref{ta:main-classes} can be formed by adding a second identical particle rotated by $90^\circ$ around the axis $z$.}
\label{omega}
\end{figure}

Here, we will find the optimal shape of  an omega particle, which, according to Section~\ref{classes}, will optimally interact with evanescent linearly polarized waves and reactive linearly polarized fields. Let us consider a canonical omega particle \cite{Engheta-omega,sochava,basic} whose shape is shown in Figure~\ref{omega}.
Similarly to the canonical chiral particle considered above, this is a connection of two small antennas: an electric dipole and a small loop, but now the particle is planar and the induced electric and magnetic moments are orthogonal.

 For simplification of derivations, we suppose that this particle is excited by a linearly polarized evanescent plane wave with the fields
 \begin{equation}
\begin{array}{l}
\displaystyle
\mh=H_x\mathbf{x}_0,\vspace*{.1cm}\\\displaystyle \me=E_y\mathbf{y}_0.
\label{eq:p}\end{array}
\end{equation}
In this case, the incident electric field is parallel to the electric dipole antenna and the magnetic field is orthogonal to the loop plane.
Because the only difference with the earlier considered canonical chiral particle is the mutual orientation of the wire and loop antennas, we can use the same expressions for the polarizabilities as in
(\ref{eq:aa}) (neglecting the electric polarization of the loop) \cite{basic}):
\begin{equation}
\begin{array}{l}
\displaystyle \alpha_{\rm ee}=-j{l^2\over \omega Z_{\rm tot}},\vspace*{.3cm}\\\displaystyle \alpha_{\rm em}=j\Omega =-
j\alpha_{\rm ee}\eta_0 {k_0S\over l},\vspace*{.3cm}\\\displaystyle
\alpha_{\rm mm}=\alpha_{\rm ee}\left(\eta_0 {k_0S\over
l}\right)^2.
\end{array}\label{eq:o}\end{equation}
The fields exponentially decay along the positive direction of the $z$-axis and the transverse wave vector is directed along $x$, with  $k_t>k_0$. The wave impedance for this evanescent wave is $Z_{\rm TE}=\omega\mu_0/\beta$, where
\begin{equation}
\beta=  n k_0= jn'' k_0=jk_0\sqrt{\displaystyle \frac{k_t^2}{k_0^2}-1}
\end{equation}
is the imaginary propagation constant of the evanescent wave, and
the imaginary value of $n=jn''$ represents the effective refractive
index. From (\ref{eq:k}) we can write the power received by
the particle as
\begin{equation}
\begin{array}{l}
\displaystyle
P=-{\o\over 2}|\me|^2{\rm Im}\left\{\aeeo+2\Omega \Im\left\{\frac{1}{Z_{\rm TE}}\right\}+\frac{\ammo}{|Z_{\rm TE}|^2}\right\}.
\label{eq:q}\end{array}
\end{equation}
Using the expressions for the polarizabilities of omega particles
(\ref{eq:o}), we are ready to find the received power at the
particle resonance:
\begin{equation}
\begin{array}{l}
\displaystyle
P={l^2\over 2}|\me|^2 \frac{1}{R_{\rm tot}}\left\{1+ 2\frac{-j\beta S}{l}
+\left(\frac{-j\beta S}{l}\right)^2\right\}.
\end{array}
\label{eq:s}\end{equation}
For the resonant particle the stretched wire length is approximately $\lambda/2$, meaning that the dipole length and the loop radius should satisfy the relation
\begin{equation}
\displaystyle l+\pi a=\frac{\lambda}{4}.
\label{eq:w}\end{equation}
Substituting in (\ref{eq:s}), we get
\begin{equation}
\begin{array}{l}
\displaystyle
P=\frac{3\pi}{\eta_0 k_0^2}|\me|^2\left\{\frac{\displaystyle \left[l+\frac{2n''}{\lambda}  \left(\frac{\lambda}{4}-l\right)^2\right]^2}{\displaystyle l^2+\frac{4}{\lambda^2}\left(\frac{\lambda}{4}-l\right)^4}\right\}.
\label{eq:x}\end{array}
\end{equation}
Equating the derivative of $P$ with respect to $l$ to zero, we can easily find the optimal values of $l$ and $a$ for which $P$ has the extremum value:
\begin{equation}
\begin{array}{l}
\displaystyle
l_{\rm opt}=\left[n''+1-\sqrt{n''(n''+2)}\right]\frac{\lambda}{4},\vspace*{.3cm}\\\displaystyle
a_{\rm opt}=\frac{-n''+\sqrt{n''(n''+2)}}{\pi}\frac{\lambda}{4}.
\label{eq:y}\end{array}
\end{equation}
Actually, in most cases the electric polarizability of omega
particles significantly depends on the electric response of the loop
part, so that the parameter $l$ in (\ref{eq:o}) is an equivalent
size which depends also on the loop diameter (see more detailed
discussion in \cite[Section~7.1.4]{basic}). However, comparing
simplified and complete analytical models as well as numerical
simulations, we have noticed that the present simplified theory
successfully predicts the optimal dimensions with the accuracy of
about 10\%.

From the last result, it is clear that for different values of the
transverse wave number (different decay rates) the optimal
dimensions are different. For the interesting special case in which
the effective refractive index equals unity in the absolute value:
$n=\pm j$ (corresponding to  $k_t=\sqrt{2}k_0$), we note that the
optimal ``scaling factor" $k_0S/l=1$, and the necessary condition
for the polarizabilities of the optimal omega particle in
(\ref{eq:o}) reads
\begin{equation}
\eta_0\alpha_{\rm ee}=\pm \Omega=\frac{1}{\eta_0}\alpha_{\rm mm}.
\label{eq:z}\end{equation}
This is again the same balance relation as for the optimal chiral particle excited by a plane wave in free space (refractive index $n=1$), telling that the optimal response corresponds to all the normalized polarizabilities being equal.

\section{General definition of optimal and balanced particles}

The above two examples of reciprocal bianisotropic particles, optimized for the maximum power extracted from the incident fields (Section~\ref{orp}), show that the optimal response is related to the condition of balanced reactions to electric and magnetic fields. According to (\ref{eq:gg}) and (\ref{eq:z}), all the normalized polarizabilities of these optimal particles are equal. Some examples of special properties of particles with balanced response to electric and magnetic excitations are known from the literature. For the simplest case of the absence of magneto-electric coupling, the balance condition $\eta_0\alpha_{\rm ee}=(1/\eta_0)\alpha_{\rm mm}$ corresponds to the Huygens particle, also called self-dual particle, e.g. \cite{Lindell_self_dual,Ziolkowski,Antti_CP,Antti_LP}. These particles have the interesting property of zero backscattering amplitude.
Special properties of small chiral particles with balances polarizabilities are also known, e.g. \cite{gent,unit}. In addition to the property of maximized or minimized strength of particle-field interactions, these particles are most suitable for realizing artificial materials with negative refractive index using the ``chiral route'' \cite{nihility0,nihility1,nihility2,nihility3,nihility4,nihility5}.
It appears that the optimal chiral particle exhibit chiral effects with the extreme strength: Optical activity in chiral media is caused by the difference in the propagation constants of the circularly polarized modes with the opposite handedness, and composites of balanced chiral particles can be engineered so that these two propagation constants differ by sign \cite{nihility1,nihility5}. In particular, it is possible to realize media with the effective refractive index $n=\pm 1$ for the two modes (in the assumption of small losses). Furthermore, the use of optimal chiral particles makes it possible to design single-layer polarization-transformation sheets, e.g. \cite{Teemu}. These examples of balanced chiral particles show that the optimization for the maximized power, extracted by the particle from the incident field, gives rise to a number of extreme electromagnetic properties of such particles and their ensembles.

In this paper we have identified the classes of optimal particles for the most general uniaxial bianisotropic particles, including nonreciprocal particles. We have seen that the requirement for optimal interaction of omega particles with exciting fields also leads to the condition of balanced polarizabilities for the special case of the effective refractive index equal to $n=\pm j$ (unity in the absolute value). We expect that not only the known chiral particles with balanced polarizabililies show extreme and useful electromagnetic properties, but also particles of all the other classes, provided that their polarizabilities are balanced. This expectation is supported by our recent studies of scattering by individual uniaxial particles and infinite periodical planar arrays \cite{Joni_single_particle,Younes_absorbers}. In these studies, we have found that conditions for zero forward or backward scattering from general uniaxial particles lead to the requirement of the balanced polarizabilities. Also, it appears that the requirements of perfect absorption, full transmission or full reflection from arrays of small particles (both reciprocal and nonreciprocal) again leads to the same condition of optimal balanced polarizabilities. Based on the above observations, we suggest to introduce the notion of the balanced particle as the particle with balanced polarizabilities for all classes of uniaxial particles, listed in Table~\ref{ta:main-classes}. This condition can be expressed in the general case as
\begin{equation}
\eta_0\alpha_{\rm ee}=\pm \alpha_{\rm em}=\pm
\frac{1}{\eta_0}\alpha_{\rm mm}. \label{eq:hh}\end{equation} The
magneto-electric parameter $\alpha_{\rm em}$ can be any of the four
possible coupling parameters: the chirality parameter $\kappa$, the
omega parameter $\Omega$, the Tellegen parameter $\chi$ and the
velocity parameter $V$. From the known extreme properties of
balanced chiral particles in the field of propagating circularly
polarized waves, we expect that the balanced particles of other
classes will show extreme responses when excited by the incident
fields of other polarizations and by reactive fields, accordingly to
the particle classes identified in Section~\ref{classes}. Finally,
we note that in many cases the balanced particles satisfy the
conditions of maximal or minimal received power, meaning that
they are also the optimal particles for interactions with incident
fields of various types, according to the classes identified above.

\section{Conclusion}

In summary, we have introduced the concept of balanced and optimal
particles and composite materials for interactions with propagating
or evanescent fields of linear and circular polarizations. The
optimal particles extract maximum power from the incident
electromagnetic fields (for a given overall size  and the resonant
frequency of the inclusion). Apparently, these particles are also
the optimal radiators of power. We have shown that the most
interesting optimal reciprocal particles have all the
polarizabilities equal (or they differ only by sign). We call such
sets of polarizabilities ``balanced''. It is expected that the use
of balanced and optimal particles especially in the design of
artificial electromagnetic materials will allow optimization of
electromagnetic performance of various devices, where the material
or layer is interacting with various types of electromagnetic waves
(antennas, absorbers, sensors, lenses, and so on).

\end{document}